# Resistivity and Magneto-Optical Evidence for Variable and Incomplete Connectivity in Dense, High Critical Current Density C-alloyed Magnesium Diboride.

B. J. Senkowicz, A. A. Polyanskii, R. J. Mungall, E. E. Hellstrom, and D. C. Larbalestier

*Applied Superconductivity Center and Department of Materials Science and Engineering, University of Wisconsin – Madison, Madison, WI 53706*

Carbon-doped magnesium diboride was fabricated from pre-reacted pure $MgB_2$ by mechanical alloying. The sample set had excellent critical current densities $J_c$(8 T, 4.2 K) ranging from 15 – 60 $kA/cm^2$, depending on composition. Magneto-optical imaging detected regions up to 0.5 mm in size which were ~100% dense with $J_c$ 2-6 times that of the matrix. Evaluation of resistivity curves using the Rowell method predicts that only 10-50% of the cross sectional area carries the normal state measurement current, suggesting that considerable increases in $J_c$ in these ~80% dense $MgB_2$ samples would be possible with complete grain and particle connectivity.

The fabrication of MgB$_2$ wires in lengths > 1 km and in stabilized multifilamentary form[1-3] demonstrates that this is a conductor technology of great interest to magnet builders.[4] But to make MgB$_2$ really viable, the critical current density $J_c$ must be raised considerably. A useful near-term benchmark would be to make $J_c$ better than NbTi. For example at 8 T and 4.2 K, this would imply an increase in $J_c$ by a factor of 3-5. This letter addresses the important obstacle of less than full connectivity that stands in the way of achieving this goal.

Most practical superconductors achieve $J_c \sim$ 10% of the depairing current density $J_d$ ~H$_c$/λ, where H$_c$ is the thermodynamic critical field (μ$_o$H$_c \sim$ 0.67 T) and λ is the London penetration depth (~140 nm), which yields $J_d$(4.2 K) ~ 4×10$^8$ A/cm$^2$. But in fact few bulk MgB$_2$ samples attain even 10$^6$ A/cm$^2$.[2-5] However, $J_c$ ~10$^7$ A/cm$^2$ has been obtained in clean, undoped MgB$_2$ films grown by hybrid physical – chemical vapor deposition (HPCVD)[6]. Such films appear to have almost perfect connectivity, but their disadvantage is that they have low H$_{c2}$ and $J_c$ decreases rapidly with increasing applied field. C-doped HPCVD films have much higher H$_{c2}$ but $J_c$ is much lower due to current blockage by non-superconducting carbon-rich phases.[6] Currently, the best[7-11] reported C-doped bulk MgB$_2$ has $J_c$(8 T, 4.2 K) in the range of 20 – 60 kA/cm$^2$. Like the C-doped films, bulk MgB$_2$ suffers from obstructed current flow, as described by Rowell.[12] This sub-optimal electrical connectivity is usually ascribed to a combination of porosity and grain boundary wetting phases[12-14], but the real reason is seldom clear. If poor connectivity is as general as the Rowell analysis predicts, in fact the intragranular $J_c$ of much bulk MgB$_2$ may already be higher than of Nb-Ti. This letter provides support for this view from a systematic study of connectivity in bulk MgB$_2$ which demonstrates that local variations of at least 3 in the current-carrying cross-section and/or the local $J_c$ value occur, even in high $J_c$ MgB$_2$.

Our sample set had compositions $Mg(B_{1-X}C_X)_2$, with X ranging from 0 to 0.1. Precursor powders of $MgB_2$, Mg, and C were mixed in stoichiometric proportions and mechanically alloyed in a SPEX mill with WC media for 600 min. Milled powders were then pressed into pellets, sealed in evacuated stainless steel tubes, and reacted in a hot isostatic press at ~0.2 GPa and 1000°C for 200 min followed by a 30 min ramp to 900°C for 300 min and then cooled at a rate of ~100°C/hr. Samples sectioned with a diamond saw were ~80% dense and lustrous when polished.

Zero-field cooled, warming $T_c$ transitions were quite sharp at low X, but beyond X ~ 0.07 broadened considerably and the higher T onset of the transition stopped decreasing. These characteristics indicate some inhomogeneity in carbon alloying beyond X ~ 0.07.[15] However, we did not observe a double transition in any of the samples, which would have indicated phase separation.

Figure 1 shows images of a polished ~110 μm thick section of the X = 0.1 sample, first by conventional light microscopy and then by magneto-optical (MO) imaging in the superconducting state. Both techniques show that the microstructure is inhomogeneous. Fig. 1a shows bright, highly reflective "islands" ranging in size from a few to >100 μm embedded in a duller, more porous matrix. Similar morphology was observed for all X ≥ 0.01. The fine microstructure is not accessible to the light microscope and is currently being studied by electron microscopy. The MO images of this same piece in Fig 1b show that polishing to thin sections did produce some large cracks, but more significantly that many of the bright islands in Fig. 1a correlate directly to important features in the MO image, as particularly noted by the three islands marked by arrows. The large MO image shows the sample after zero-field cooling to 10 K and then after applying 80 mT, which is not high enough to penetrate the whole sample.

The small dark spots visible in the large MO image correspond to dense islands. The dark MO contrast of the islands indicates that they preferentially exclude flux compared to the matrix around them. The higher field (120 mT) inset MO image in Fig. 1b shows field penetrating the whole sample. The flux gradient was measured along line C-D and found to be at least 3 times steeper in the island compared to the matrix. Other islands (including those marked by the arrows) had 2-6 times larger gradients than that of the matrix, indicating a proportional variation of $J_c$.

Using scanning electron microscopy we found substantial porosity in the matrix, but almost full density in the islands, as shown in Fig 2. The near absence of porosity in Fig. 2b is a rare example of complete sintering and densification in C-doped $MgB_2$ powder without adding Mg from outside the pellet.

The Rowell analysis[12] provides a means to estimate the fraction ($A_F$) of the sample cross section that carries the current in a normal-state resistivity measurement ($\rho(T)$). Assuming that fully connected pure $MgB_2$ has a well defined scattering that yields a characteristic $\Delta\rho = \rho(300\ K) - \rho(>T_c)$ regardless of fabrication process, it is possible to separate the effects of electron scattering from connectivity on the normal state $\rho(T)$ curve and thus calculate the true resistivity of the connected portion of the sample, which Rowell called $\rho(0)$ and we call the adjusted resistivity ($\rho_A$). However, in this study we also need to account for the extra scattering produced by C to make realistic estimates of $A_F$. To do this we used $\Delta\rho$ data taken on dense C-doped filaments[16-17].

Figure 3 gives measured $\rho(40K)$, $\rho_A(40\ K)$, and $A_F$ values for the sample set. Resistivity values vary over about two orders of magnitude and have a large scatter, but do show a general increase with increasing X. Despite the wide scatter in the measured $\rho(40\ K)$, the adjusted true resistivity $\rho_A$(40 K) is a smooth increasing function of X, indicating that sample-to-sample variation in connectivity is well accounted for by the Rowell analysis. Calculated $A_F$ values are widely scattered and range from ~0.14 to ~0.50, indicating that electrical connectivity is significantly impeded, even in these ~80% dense samples. In fact the modeling of current flow in superconductors shows that physical obstructions to the current flow are significantly larger than in the normal state due to the highly non-linear voltage-current characteristics, making it clear that the local $J_c$ will be higher than simply dividing the critical current by $A_F$.[18] It should be stressed that although high density is beneficial to electrical connectivity, even a fully dense sample could have poor connectivity due to wetting grain boundary phases.

Figure 4 shows whole-sample $J_c$ values calculated from magnetization loops taken in a vibrating sample magnetometer (VSM) on rectangular section bars using the Bean model formula $M=J_c d((3b-d)/d^2)$ where b and d are the sample dimensions perpendicular to the applied field. Uncertainty in b and d was probably ~10%. For this sample (X = 0.1), $J_c$(0T and 8 T at 4.2 K) was ~200 and ~15 kA/cm$^2$, respectively, while for other compositions $J_c$(8T, 4.2K) was as high as ~60 kA/cm$^2$, amongst the highest $J_c$ values found in bulk samples. Consistent with our earlier comment, a $J_c$ (0T, 4.2K) of ~200 kA/cm$^2$ is rather low, only 0.005 $J_d$.

From the flux gradients observed by MO along line C-D in Fig. 1b, we calculated the local $J_c$ of the matrix and an island.[19] The inset to Fig. 4 compares $J_c$(T) of the matrix estimated from the MO images with the VSM data. Their close agreement indicates that the VSM-derived $J_c$ values are determined by the whole-sample matrix currents. We found $J_c$ of the islands to be 2 to 6 times higher than that of the matrix, suggesting that full densification is an important route to raising $J_c$ in practical forms of alloyed MgB$_2$. At present their effect on the whole sample $J_c$ is limited because the islands are disconnected from each other.

In summary, these dense, hot isostatic pressed high $J_c$ samples were found to be much denser than normal samples[20] but still physically and electrically inhomogeneous. Although possessing an already high $J_c$ for bulk samples, significantly higher $J_c$ islands of almost full density surrounded by a lower $J_c$, ~80% dense matrix were observed. Rowell connectivity analysis predicted that the X=0.1 sample had an effective cross section of only ~0.3 for normal state currents flowing across the matrix. This reduced $A_F$ factor suggests at least a factor of 3 increase in local flux pinning current density, a result which is broadly consistent with the 2 to 6 times higher local current density of fully dense regions indicated by MO flux gradient measurements. Thus it is clear that one path to obtaining higher current densities in $MgB_2$ that are comparable to Nb-Ti is to fully densify the material. Further densification experiments are underway to check this conjecture.

BJS was supported by the Fusion Energy Sciences Fellowship Program, administered by Oak Ridge Institute for Science and Education under a contract between the U.S. Department of Energy and the Oak Ridge Associated Universities. This work was supported by the NSF – FRG on $MgB_2$, and by DOE – Understanding and Development of High Field Superconductors for Fusion - DE-FG02-86ER52131. The authors thank their Madison colleagues J. Jiang, W. Starch, A. Squitieri, and J. Mantei and visiting scientists in Madison, A. Matsumoto of NIMS, and A. Yamamoto of the University of Tokyo for discussions.

Figure 1 – a) Optical image of a polished thin section with X = 0.1.  b) Magneto – optical (MO) images of the same sample.  Large MO image was zero-field cooled to 10 K , then 80 mT was applied.  Cracks are marked with "X"  Inset MO image was zero-field cooled to 10 K before 120 mT was applied.

Figure 2 – Secondary electron micrographs of the X = 0.1 sample fracture surface.  a) Representative porous matrix.  The brightness is due to the high density of edges within the microstructure.  b) Representative dense island.

Figure 3 – Measured $\rho$(40 K), calculated $\rho_A$(40 K) and calculated $A_F$ as a function of nominal composition using $\Delta\rho$ data taken from dense, low-resistivity, carbon doped filaments.[17]

Figure 4 – VSM $J_c$ as a function of applied field at a variety of temperatures for X = 0.1. Inset shows $J_c$ as a function of temperature for X = 0.1 at low field as measured by VSM (open squares) and as reconstructed from flux penetration gradients across the *matrix area* along line C-D in Fig. 1b measured by MO with H < 200 mT.

Figure 1

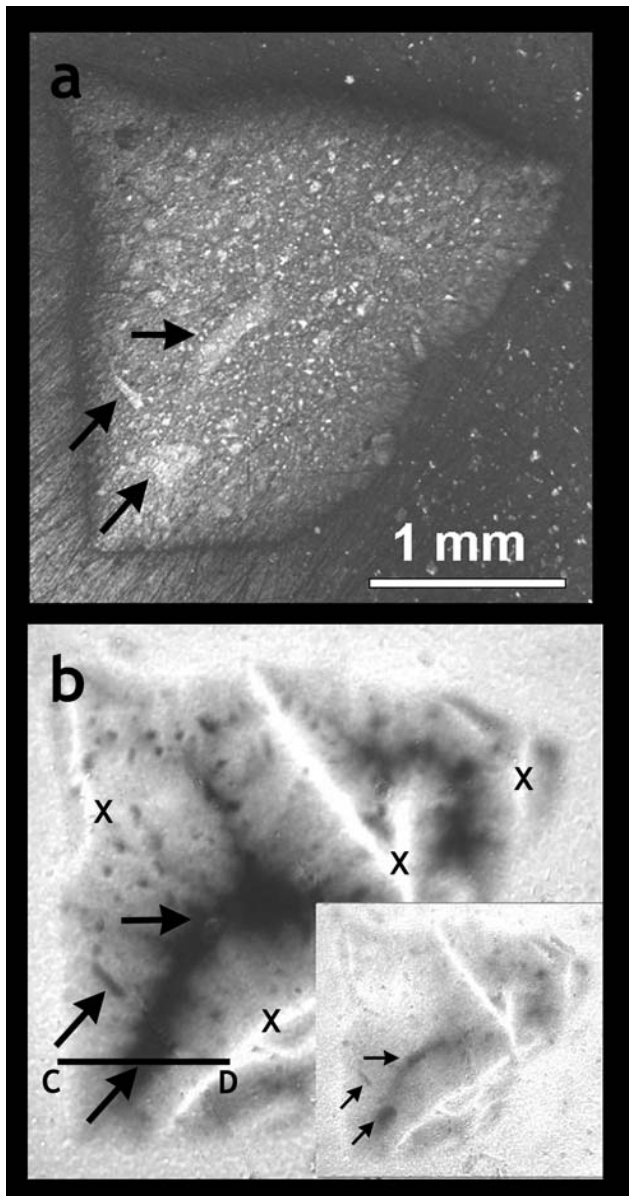

Figure 1

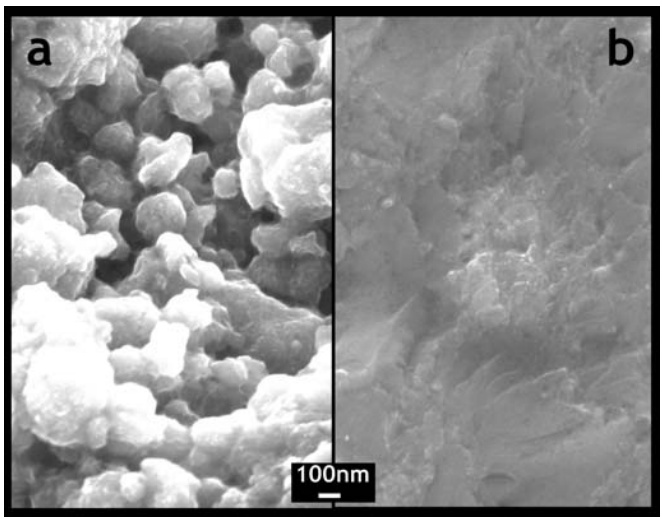

Figure 3

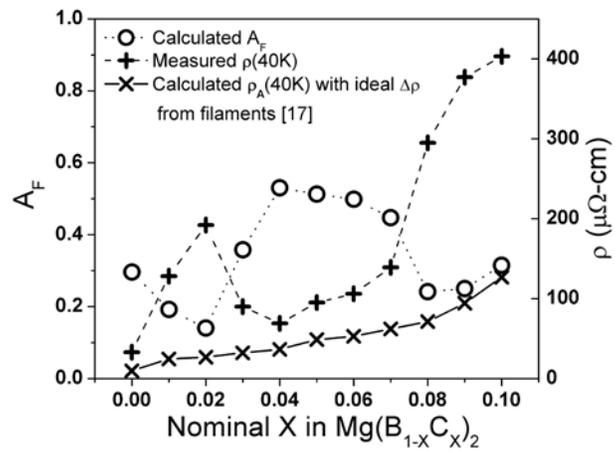

Figure 4

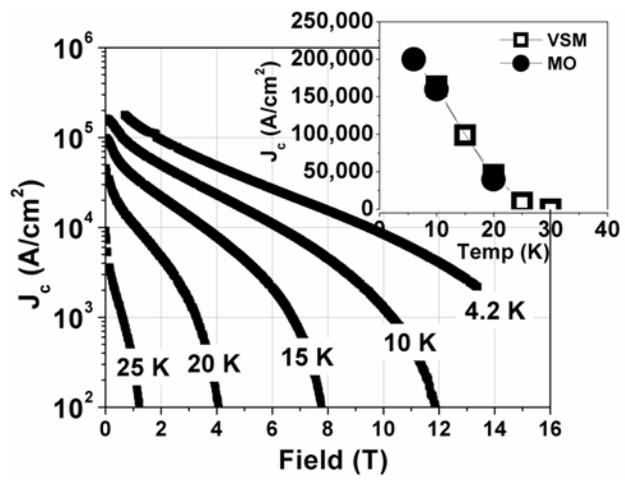